\documentstyle[emulateapj,onecolfloat,psfig]{article}   

\newcommand{\gal}{{\rm gal}}

\newcommand{\lin}{{\rm lin}}

\newlength{\tskip}
\newlength{\colwidth}

\newcommand{\beq}{\begin{equation}}
\newcommand{\eeq}{\end{equation}}
\newcommand{\beqa}{\begin{eqnarray}}
\newcommand{\eeqa}{\end{eqnarray}}

\begin{document}
\twocolumn[   

\title{Second Moment of Halo Occupation Number}
\author{Asantha Cooray\altaffilmark{1}}
\affil{
Division of Physics, Mathematics and Astronomy, California
Institute of Technology, Pasadena, CA 91125.\\
E-mail: asante@caltech.edu}

\begin{abstract}
The halo approach to large scale structure provides a physically motivated 
model to understand clustering properties of galaxies. An important aspect of the halo model
involves a description on how galaxies populate dark matter halos or what is now called the
halo occupation distribution.
We discuss a way in which clustering information, especially in the non-linear regime,
can be used to determine moments of this halo occupation number. 
We invert the non-linear part of the real space power spectrum from the 
PSCz survey  to determine the second moment of the halo occupation distribution in a model
independent manner.  The precise measurement 
of higher order correlations can eventually be used  to determine successive higher 
order moments of this distribution.
\end{abstract}

]

\altaffiltext{1}{Sherman Fairchild Senior Research Fellow} 

\section{Introduction}

The halo approach to large scale structure has now become a useful tool to study and
understand  clustering properties of dark matter and a number of tracers including galaxies
(see \cite{CooShe02} 2002 for a recent review). This approach replaces the complex
distribution of dark matter with a collection of collapsed dark matter halos. Thus, necessary inputs
for a halo based model include properties of this dark matter halo population, such as its
mass function and the spatial profile of dark matter within each halo 
(\cite{Sel00} 2000; \cite{MaFry00} 2000; \cite{Scoetal00} 2000; \cite{Cooetal00} 2000).

In order to describe clustering aspects beyond dark matter, it is necessary that one
understands how the tracer property is related to the dark matter distribution in each halo.
In the case of galaxies, an important input is a description on how galaxies
populate halos. This is usually achieved by the so-called halo occupation number
where one describes the mean number of galaxies in dark matter halos as a function of mass and its
higher order moments (\cite{Sel00} 2000; \cite{PeaSmi00} 2000; \cite{BerWei02} 2002).
For the two-point correlation function, one requires information up to the
second moment of the halo occupation distribution, while higher order correlations
successively depend on  increasing moments.

The halo occupation distribution has been widely discussed in the literature in terms of semi-analytical
models of galaxy formation (e.g., Benson et al. 2001; Somerville et al. 2001). With the advent of
a well defined halo approach to clustering, observational constraints have also begun to appear 
(e.g., \cite{Scoetal00} 2000; \cite{MouSom02} 2002). 
While expectations for constraints on the halo
occupation distribution from current wide-field galaxy surveys, such as the Sloan Digital Sky Survey, 
are high (\cite{BerWei02} 2002;
\cite{Scr02} 2002), these are, however, all considered in a model dependent manner.

Though descriptions on the halo occupation number based on a specific model is useful, it is probably more
useful to consider model independent constraints.  
In this {\it Letter}, we consider such an approach and suggest that clustering
information, especially in the non-linear regime, can be used for a reconstruction of
various moments of the halo occupation number. We discuss a possible inversion for this purpose
and use results on the non-linear power spectrum from the PSCz redshift survey (Saunders et al. 2000) by 
Hamilton \& Tegmark (2002; see also, Hamilton et al. 2000) 
to provide a first estimate of the second moment of the halo occupation number.
We discuss both strengths and limitations of our approach and provide a comparison to
model based descriptions of the halo occupation number.

We provide a general discussion of our method in the next section.
when illustrating results in \S~3, we take a flat $\Lambda$CDM cosmology 
with parameters $\Omega_c = 0.3$, $\Omega_b=0.05$, 
$\Omega_\Lambda=0.65$, $h=0.65$, $n=1$, $\delta_H=4.2\times 10^{-5}$.

\section{Halo Approach to Galaxy Clustering}

The halo approach to galaxy clustering assumes that 
large scale structure can be described by a distribution of dark matter halos while
galaxies themselves form within these halos. At the two-point
level of correlations, the contribution can be written as a sum of correlations of galaxies that occupy
two different halos and the correlation of galaxies within a single halo.
These two terms are generally identified in the literature as 2- and 1-halo terms, respectively.
Under the assumption that halos trace the linear density field, the 2-halo term is
proportional to the linear power spectrum  with a bias factor that can be calculated from
analytical arguments (\cite{Moetal97} 1997). 
The mass function, such as the Press-Schechter (PS; \cite{PreSch74} 1974),  
and the distribution of dark matter in halos, such as the NFW profile of Navarro et al. (1996),
are known from either analytical or numerical techniques.

\begin{figure}[t]
\centerline{\psfig{file=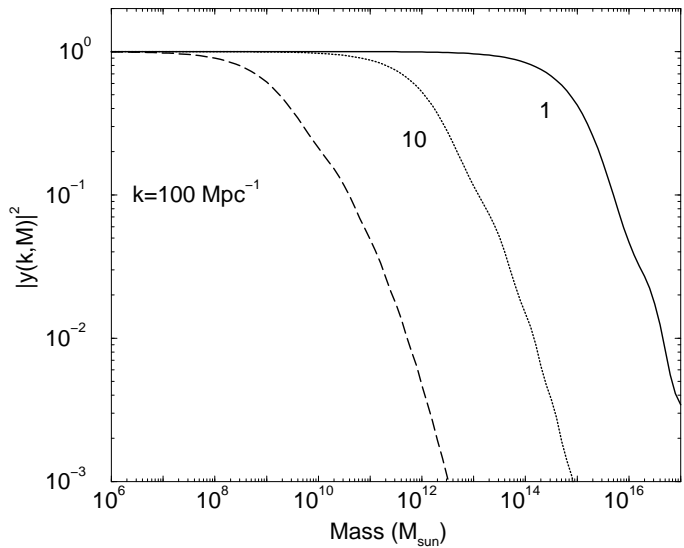,width=3.5in,angle=-90}}
\caption{$|y(k|m)|^2$ as a function of mass for three different values of $k$, as written on the plot,
at redshift of zero. We assume dark matter profiles under the description of Navarro et al. (NFW; 1996).}
\label{fig:ym}
\end{figure}

Assuming a halo population with a mass function given by $n(m)$, and that galaxies trace the
dark matter in each halo randomly, we can write the total contribution to the power spectrum of galaxies as
\begin{eqnarray}
 P_\gal(k) &=& P^{1h}_{\gal}(k) + P^{2h}_{\gal}(k)\, ,
               \qquad\qquad{\rm where}\nonumber\\
 P^{1h}_{\gal}(k) &=& \int dm \, n(m) \,
                    \frac{\left< N_\gal(N_\gal-1)|m\right>}{\bar{n}_\gal^2}\,
                    |y(k|m)|^2 \,,\nonumber\\
P^{2h}_{\gal}(k) &=& P^\lin(k) \left[ \int dm\, n(m)\, b_1(m)\,
\frac{\left< N_\gal|m\right>}{\bar{n}_\gal}\, y(k|m)\right]^2 .
\label{eqn:gal}
\end{eqnarray}
Here, $y(k|m)$ represents the three-dimensional Fourier transform of the dark matter profile. 
 In general, the 1-halo term dominates the total contribution to the power spectrum
 in the non-linear regime while the 2-halo term captures the large scale correlations in the linear regime.  

In equation~\ref{eqn:gal}, $\left< N_\gal|m\right>$ and $\left< N_\gal(N_\gal-1)|m\right>$ are the first and
second moments of the galaxy occupation distribution, respectively, $\bar{n}_\gal$ is 
mean number density of galaxies,
\begin{equation}
\bar{n}_\gal = \int dm \, n(m)\, \left< N_\gal|m \right> \,,
\label{eqn:barngal}
\end{equation}
$b_1(m)$ is the first order halo bias (Mo et al. 1997), and $P^\lin(k)$ is the linear power
spectrum of the density field. On large scales where the two-halo term dominates and
$y(k|m)\to 1$, the galaxy power spectrum simplifies to
\begin{equation}
P_{\gal}(k) \approx b_\gal^2\, P^\lin(k),
\end{equation}
where
\begin{equation}
b_\gal = \int dm\, n(m)\, b_1(m)\,
         \frac{\left< N_\gal|m\right>}{\bar{n}_\gal}\, ,
\end{equation}
represents the large scale constant bias factor of the galaxy population. 
Though we have not considered, in detail,
there are slight modifications to equation~(1) especially if one is interested in accounting for
possibilities such as a single galaxy always form at the center of each dark matter halo and the fact that
galaxies may not trace the dark matter perfectly. 

At non-linear scales, $y(k|m) \ll 1$, and the 1-halo term dominates. Since $y(k|m)$ is no longer constant,
if $y(k|m)$ is assumed a priori from a certain dark matter profile,
we can consider  a possible inversion of the power spectrum measurements to reconstruct
$f(m) \equiv n(m) \left< N_\gal(N_\gal-1)|m\right>/\bar{n}_\gal^2$.
 With information related to the halo mass function and mean density of galaxies, 
which comes directly from data, one can obtain a model independent estimate of
$\left< N_\gal(N_\gal-1)|m\right>$.  Note that in the large scale regime, because of the scale-independent
constant bias, the equation is non-invertible. Therefore, clustering information cannot be used reconstruct the
mean of the halo occupation number appropriately.

Though our use of outside knowledge on $y(k|m)$  and $n(m)$ 
may make the extraction of the second moment model dependent, one should note that these
are effectively properties of the dark matter and are determined well from N-body numerical simulations.
The fact that  we have a potential method to estimate 
information on the galaxy side, mainly details on the halo occupation number
 without resorting to any models, 
is extremely important. This is the main result of this paper. We now consider the possibility for
an inversion and an application of  our suggestion to galaxy clustering data at non-linear
scales from the PSCz survey (Saunders et al. 2000).

\section{Inversion}

In order to invert the non-linear power spectrum to estimate information
related to second moment of the halo occupation number,
we follow standard approaches in the literature related to inversions associated with
large scale structure clustering (e.g., Dodelson et al 2001; Cooray 2002). 
These inversions are usually considered to recover three-dimensional
clustering information  from two-dimensional angular clustering data.
The inversion problem here is similar: Given estimates of $P_\gal^{2h}(k)$,
and information related $y(k|m)$, we would like to estimate the function $f(m)$ defined earlier.

\begin{figure}[t]
\centerline{\psfig{file=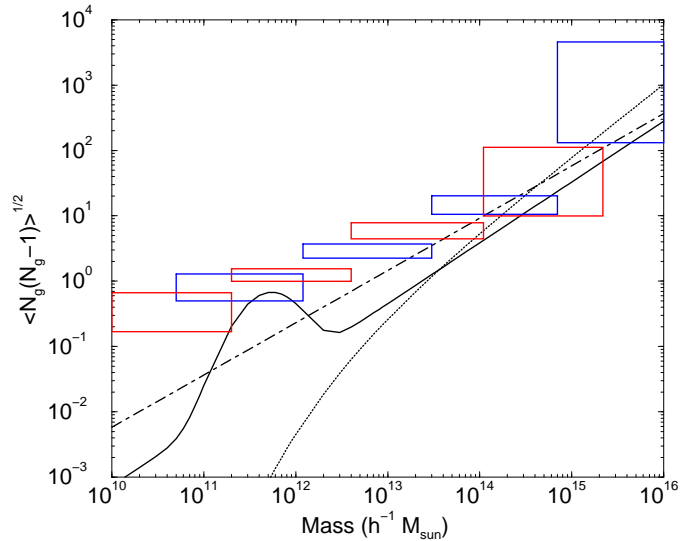,width=3.5in,angle=-90}}
\caption{The second moment of the halo occupation number, plotted as 
$\sqrt{\left< N_\gal(N_\gal-1)|m\right>}$, as a function of halo mass $m$.
The two color error boxes show two indepdent sets of estimates with bins of one set shifted 
relative to those of the other. For comparison, we show three estimates on the second moment of the
halo occupation: The blue- (solid) and red-galaxy (dotted) model of Scranton (2002) and
the best fit blue galaxy model of Scoccimarro et al. (2000) based on APM 
data (dot-dashed line) with normalization following Sheth \& Diaferio (2001).}
\label{fig:nm}
\end{figure}

We can write the associated inversion equation  as
\begin{equation}
\vec{P} = {\bf Y} \vec{M} + \vec{n} \, ,
\label{eq:vector}
\end{equation}
where,  $\vec{P}$ is a vector containing the data related to  $P_\gal^{2h}(k)$
with an associated noise vector $\vec{n}$, and ${\bf Y}$ is a matrix containing kernels at each $k_i$, 
where the non-linear power spectrum  is measured, and at each $m_i$, for which galaxy weighted mass function information 
is desired.  The inversion problem stated in terms of this equation involves estimating $\vec{M}$ given
other vectors and the matrix ${\bf Y}$. Note that the matrix ${\bf Y}$ differs from kernels defined by $|y(k|m)|^2$
due to an additional factor of $dm$. By appropriately renormalizing equation~(\ref{eq:vector}) with noise,
following Dodelson et al. (2001), we can consider the minimum variance estimate of $f(m)$. We refer the reader
to Dodelson et al. (2001) for full details of the inversion. Additional discussions on
related inversion techniques are available in \cite{DodGaz00} (2000), \cite{EisZal01} (2001) and references therein.

In figure~1, we show $|y(k|m)|^2$ as a function of halo mass, $m$, for
three different values of $k$ and at a redshift of zero. The behavior of these kernels are important for
the inversion since they determine how likely the inversion will be stable and to what extent information can be
extracted on $f(m)$. The effective width of $|y(k|m)|^2$ increases with decreasing $k$ similar to the 
behavior one observes from kernels associated the inversion of a simple galaxy
correlation function. The kernels in the latter case involves a $J_0$, the zeroth order Bessel function, which
when plotted looks similar in shape to associated kernels here, except for the ringing part associated with
$J_0$ functions that oscillate between positive and negative values.

The general behavior of present kernels,  as a function of $k$ and $m$, is 
also consistent with how the halo model contributes to the non-linear power spectrum:
at small physical scales, contributions come from the low mass halos while at large physical scales
or small $k$ values, contributions to the power spectrum come from the whole mass range. The overlap
of kernels with decreasing $k$ 
suggests that any estimates of the non-linear power spectrum is likely to be highly correlated. This
is, in fact, true when one considers the covariance resulting from the four-point correlations function or
the trispectrum (e.g., \cite{CooHu01} 2001).
Thus, any measurement of
the non-linear power spectrum should be considered with its associated covariance matrix and correlations should be
properly accounted when inverting the power spectrum to determine any physical property, such as $f(m)$ defined above.

In terms of published analyses of clustering, this information, unfortunately, is not fully available to us 
from the literature. The best published estimate of
the non-linear galaxy power spectrum, so far, comes from the PSCz survey by Hamilton \& Tegmark (2002).
Given lack of knowledge on its covariance, as advised by one of the authors, we used the ``prewhitened'' power
spectrum estimated by the same authors to carry out an inversion as part of this study. 
It is suggested in Hamilton \& Tegmark (2002) that estimates of the
prewhitened power spectrum are decorrelated and we used estimates and their errors as part of this
inversion without accounting for the correlations. 

For the purpose of this inversion, we define
the non-linear part as the power spectrum from $k > 1$ h Mpc$^{-1}$. At small $k$ values, we expect a fractional
contribution from the 2-halo term, or the correlated part between galaxies in different halos. This contribution,
however, decreases rapidly with increasing value for $k$. We use power spectrum 
estimates out to $k \sim 300$ h Mpc$^{-1}$, though,
estimates beyond $\sim$ 100 h Mpc$^{-1}$  from the PSCz survey are noise dominated. 
Following the redshift distribution of
PSCz as measured by Saunders et al. (2000), we assume a redshift of 0.028 for the full non-linear power spectrum.
It is likely that estimates of the non-linear power are redshift dependent (Scranton \& Dodelson 2000), 
however, we have no useful information
to account for such variations. 

When converting $f(m)$ estimates to the second moment of the halo occupation
number, we require information on the number density of
PSCz galaxies at the mean redshift of the estimated non-linear power spectrum. 
We obtained this information using the luminosity function of PSCz galaxies as calculated by
Seaborne et al. (1999; see also Saunders et al. 1990), but we have not attempted to account for any
variations on the density from the original PSCz catalog to the one utilized in estimating the power
spectrum by Hamilton \& Tegmark (2000).  We expect any changes on this aspect, however, if any at all, to be minor.
Also, in order to convert $f(m)$ to the second moment, we assume the halo mass 
function is described by \cite{PreSch74} (1974) theory. Again, we make no attempt to incorporate any uncertainties in
the mass function.

We summarize our results on the second moment of the halo occupation distribution in figure~2. Here, we plot
$\left< N_\gal(N_\gal-1)|m\right>^{1/2}$ as a function of the halo mass $m$. We present two separate sets of estimates
of this quantity with bands, on mass axis, of the second set shifted relative to the bands of the
first set. This shifting demonstrates the robustness of the inversion and follows from the now well known 
analysis technique introduced by the cosmic microwave background experimentalists. Note that our estimates are
likely to be contaminated by assumptions used in the analysis. One of the main drawbacks is the lack of an accounting
of full covariance, or associated correlations, between power spectrum measurements. Another is the assumption
that galaxies are randomly distributed in each halo with no preference to be at the halo center.
The latter assumption is likely to affect the estimates at the low mass end especially when 
$\left< N_\gal(N_\gal-1)|m\right>^{1/2}$ goes below 1. Since we find this to be the case only in our first bins,
we have not attempted to correct for this any further. 

The second moment of the halo occupation number estimated here should be considered as the value
at the mean redshift of the PSCz catalog. The estimates clearly show the power-law behavior 
of the halo occupation number.
If the distribution can be described by a Poisson distribution, then we can write 
$\left< N_\gal(N_\gal-1)|m\right>^{1/2} = \left< N_\gal|m\right>$. For comparison, we also plot
predictions for the mean number, $\left< N_\gal|m\right>$, following various analyses in the literature.
These include model descriptions by Scranton (2002) using results from semi-analytical models of
galaxy formation and from Scoccimarro et al. (2000) from power law fits to the APM clustering data.
The latter can be described as $\left< N_\gal|m\right> = A(m/m_0) ^{0.8}$. We normalize this curve
using  $m_0=4 \times 10^{12} h^{-1} M_{\sun}$ and $A=0.7$ following Sheth \& Diaferio (2001).

At intermediate mass ranges and below,  
estimated band powers generally fall above these descriptions 
suggesting that $\left< N_\gal(N_\gal-1)|m\right>^{1/2}$ may follow as $\alpha(m) \left< N_\gal|m\right>$.
There are descriptive models available in the literature 
for $\alpha(m)$ (Scoccimarro et al. 2000; Scranton 2002). 
These, however, indicate $\alpha(m) < 1$ such that the halo occupation number is sub-Poissonian at the low
mass end. This is contrary to estimates we have obtained.
In general, Scranton's blue galaxy model is better consistent with our estimates, though, a priori, there is no
reason to believe why PSCz galaxies should be described by such a model.

Assuming the second moment can be described by a simple power law such that $\left< N_\gal(N_\gal-1)|m\right>^{1/2}=(m/m_0)^p$, we fitted our band power estimates to constrain $(m_0,p)$. When estimating the
likelihood, we use the full covariance matrix of our estimates of $\left< N_\gal(N_\gal-1)|m\right>^{1/2}$. 
We show 1-, 2- and 3-sigma contours in figure~3. In general, our estimates are consistent with a wide range of values of
$m_0$, from $10^{10}$ to $10^{13}$, and $p$, from 0.2 to 1, at the 3-sigma level while power laws with slopes 
less than 0.8 are consistent at the 2-sigma level. 

\begin{figure}[t]
\centerline{\psfig{file=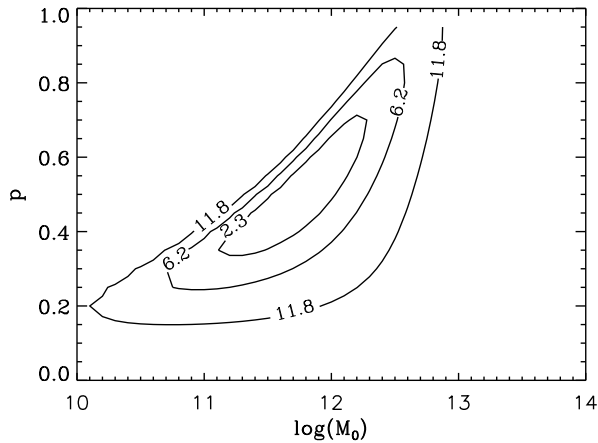,width=3.5in,angle=90}}
\caption{Constraints on the slope, $p$, and normalization, $m_0$, of the halo occupation number from the PSCz
data. We show 1-,2- and 3-sigma contours levels which are labeled as $\Delta \chi^2=$2.3, 6.2 and 11.8, respectively.}
\end{figure}

The present technique can be
extended for higher order correlations as well. Note that in the non-linear regime, under the halo model,
the p-point power spectrum, especially in the case of equal length configurations with $k_i=k$, can be written as
\begin{eqnarray}
&&T_p^{1h}(k) = \\
&& \int dm \, n(m) \,
                    \frac{\left< N_\gal(N_\gal-1)...(N_\gal-n)|m\right>}{\bar{n}_\gal^p}\,
                    |y(k|m)|^p \,, \nonumber
\end{eqnarray}
where $n=p-1$. 
The inversion technique is effectively similar and can be used to
determine related higher order moments of the
galaxy halo occupation number. And, once again, such
an approach will require higher order p-point power estimates down to the highly non-linear regime and, 
more importantly, a proper accounting of associated errors.
As always, we look forward to the day such a study can be carried out with
observational measurements from ongoing and upcoming wide-field surveys such as the Sloan and many others.

\section{Summary}

The halo approach to large scale structure provides a physically motivated 
technique to study clustering properties of galaxies (Cooray \& Sheth 2002 and references therein). 
A necessary, and an important, ingredient  for a halo based clustering calculation involve a description of how 
galaxies populate dark matter halos or the so-called halo occupation number.
We have raised the possibility for a model independent study on moments of the halo occupation number using
an inversion of the non-linear clustering power, and p-point, 
spectrum measurements. We have considered an application of this suggestion utilizing the 
PSCz power spectrum estimated by  Hamilton \& Tegmark (2002).
Our estimates on the second moment of the halo occupation number
are  consistent with power law models  over five decades in mass and 
with certain model descriptions in the literature.
With expected increase in measurements of the clustering in the non-linear regime, and associated
measurements of covariance, we expect analysis like the one suggested here will eventually
make it possible for a detailed understanding of the nature of galaxy occupation in halos.

\smallskip
{\it Acknowledgments:} This research was supported at Caltech by a senior research fellowship from
the  Sherman Fairchild foundation and a DOE grant (DE-FG03-92-ER40701). 
We thank Ryan Scranton and Andrew Hamilton for useful suggestions and appreciate
the quite atmosphere at the Aspen Center for Physics where this work was initiated.


\begin{thebibliography}{99}
\frenchspacing

\bibitem[Berlind \& Weinberg]{BerWei02}
	Berlind, A. B. \& Weinberg, D. H. 2002, ApJ in press (astro-ph/0109001).

\bibitem[Benson et al]{Benetal00}
        Benson, A. J., Pearce, F. R., Frenck, C. S., Baugh, C. M., Jenkins, A.
2001, MNRAS, 320, 261.

\bibitem[Cooray \& Hu]{CooHu01}
        Cooray, A., Hu, W. 2001, ApJ, 554, 56.

\bibitem[Cooray \& Sheth]{CooShe02} Cooray, A. \& Sheth, R. K. 2002, to appear in Physics Reports.

\bibitem[Cooray]{Coo02} Cooray, A.  2002, MNRAS submitted (astro-ph/0206068).

\bibitem[Cooray et al]{Cooetal00}
        Cooray, A., Hu, W., Miralda-Escud\'e, J. 2000b, ApJ 536, L9

\bibitem[Dodelson \& Gazta\~naga]{DodGaz00}
Dodelson, S. \& Gazta\~naga, E. 2000, MNRAS, 312, 774
 
\bibitem[Dodelson et al.]{Dodelat01}
        Dodelson, S., Narayanan, V. K., Tegmark, M. et al. 2001, ApJ in press
(astro-ph/0107421).

\bibitem[Eisenstein \& Zaldarriaga]{EisZal01}
        Eisenstein, D. J. \& Zaldarriaga, M. 2001, ApJ, 546, 2.

\bibitem[Hamilton \& Tegmark]{HamTeg02}
	Hamilton, A. J. S. and Tegmark, M. 2002, MNRAS, 330, 506.

\bibitem[Hamilton et al]{Hametal00}
		Hamilton, A. J. S., Tegmark, M. \& Padmanabhan, N. 2000, MNRAS, 317, L23

\bibitem[Ma \& Fry]{MaFry00}
        Ma, C.-P., Fry, J. N. 2000, ApJ, 543, 503.

\bibitem[Mo et al]{Moetal97}
        Mo, H. J., Jing, Y. P., White, S. D. M. 1997, MNRAS, 284, 189.

\bibitem[Moustakas \& Somerville]{MouSom02}
	Moustakas, L. A. \& Somerville, R. S. 2002, ApJ submitted (astro-ph/0110584).

\bibitem[Navarro et al]{Navetal96}
        Navarro, J., Frenk, C., White, S. D. M., 1996, ApJ, 462,
563.

\bibitem[Peacock \& Smith]{PeaSmi00}
	Peacock, J. A. \& Smith, R. E. 2000, MNRAS, 318, 1144.

\bibitem[Press \& Schechter]{PreSch74}
        Press, W. H., Schechter, P. 1974, ApJ, 187, 425.

\bibitem[Saunders et al]{Sauetal00}
	Saunders, W., Sutherland, W. J., Maddox, S. J. et al. 2000, MNRAS, 317, 55.

\bibitem[Saunders et al]{Sauetal90}
	Saunders, W., Rowan-Robinson, M., Lawrence, A. et al. 1990, MNRAS, 242, 318.

\bibitem[Scoccimarro et al.]{Scoetal00}
        Scoccimarro, R., Sheth, R., Hui, L. \& Jain, B. 2000, ApJ, 546, 20.

\bibitem[Scranton]{Scr02}
Scranton, R. 2002, ApJ submitted (astro-ph/0205517).

\bibitem[Scranton \& Dodelson]{ScrDod00}
Scranton, R. \& Dodelson, S. 2000, MNRAS submitted (astro-ph/0003034).

\bibitem[Seaborne et al]{Seaetal99}
	Seaborne, M. D., Sutherland, W., Tadros, H. et al. 1999, MNRAS, 309, 89.

\bibitem[Seljak]{Sel00}
        Seljak, U. 2000, MNRAS, 318, 203.

\bibitem[Sheth \& Diaferio]{SheDia01}
        Sheth, R. K., Diaferio, A. 2001, MNRAS, 322, 901

\bibitem[Somerville et al]{Sometal01}
	Somerville, R. S., Lemson, G., Sigad, Y., Dekel, A., Kauffmann, G., White, S. D. M. 2001, MNRAS, 320, 289.



 \end{thebibliography}
\end{document}